# Highly efficient narrowband terahertz generation driven by a two-spectral-line laser in PPLN

H. T. Olgun[1,+], W. Tian[1,2,+], G. Cirmi[1,3,+], K. Ravi, C. Rentschler[1],
H. Çankaya[1,3,4], M. Pergament[1], M. Hemmer[1], N. H. Matlis[1] AND
F. X. Kärtner[1,3,4,*]

[1]*Deutsches Elektronen-Synchrotron (DESY), Center for Free-Electron Laser Science, Notkestrasse 85, 22607 Hamburg, Germany*
[2]*School of Physics and Optoelectronic Engineering, Xidian University, Xi'an 710071, China*
[3]*The Hamburg Centre for Ultrafast Imaging, University of Hamburg, Luruper Chaussee 149, 22761 Hamburg, Germany*
[4]*Physics Department, University of Hamburg, Luruper Chaussee 149, 22761 Hamburg, Germany*
[+]*these authors contributed equally*
[*]*franz.kaertner@cfel.de*

**We demonstrate record conversion efficiencies of up to ~0.9% for narrowband (<1% relative bandwidth) THz generation by strongly-cascaded difference frequency generation. These results are achieved using a novel laser source, customized for high efficiencies, with two narrow spectral lines of variable separation and pulse duration (≥250 ps). THz generation in 5% MgO doped PPLN crystals of varying poling period was explored at cryogenic and room temperature operation as well as different crystal lengths. This work addresses an increasing demand for high-field THz pulses which has, up to now, been largely limited by low optical-to-THz conversion efficiencies. ©2021 Optica**

## 1. INTRODUCTION

Narrowband, high-field terahertz (THz) radiation has applications in a wide range of fields, including imaging [1], linear and nonlinear THz spectroscopy [2], tuned excitation of material transitions [3], and more recently in powering novel accelerators with the potential to revolutionize compact electron sources and related research areas. Until recently, research in nonlinear optical generation of THz has focused on development of broadband, single-cycle sources [4,5], due to their ability to provide ultra-high peak fields which are highly sought for controlling material properties. As a result, conversion efficiencies approaching 1% have been demonstrated [6]. Achieving comparable peak fields with the long durations associated with narrowband THz pulses requires significantly more energy, so techniques for developing these sources have therefore not seen as much development. Emerging applications, such as THz-driven electron acceleration, however, are increasingly calling for high-field pulses with high spectral purity, spurring multiple advancements which have driven efficiencies from the $10^{-5}$ to the $10^{-3}$ range and THz pulse energies from the nJ to the mJ range [7-10]. Despite these advances, low conversion efficiencies remain a limiting factor for achieving the pulse energies needed by many applications. For example, THz-driven electron accelerators require pulses with tens of millijoules of highly monochromatic radiation in the 0.1 – 1 THz frequency band with peak power in the 100 MW range and focused fields of several hundred MV/m [4]. These numbers are currently about two orders of magnitude beyond the state of the art. This low figure can be attributed to the large discrepancy between optical and THz photon energies which yields sub-percent energy conversion even for 100% photon conversion. Simulations have shown, however, that multiple-percent conversion efficiencies, far beyond the Manley-Rowe limit, are achievable if the optical pulses are tailored to promote cascading of the nonlinear interaction [11], and hence, multiple conversion processes per photon [12-13].

Key among the properties are the optical spectrum, which should be composed of a series of narrow lines separated by the THz frequency, and the pulse duration, which should match the length of the crystal. Although 2-line, mW sources are commercially available for THz generation in the continuous wave regime (CW) [14], high-energy pulsed sources do not currently exist. Here, we develop a multi-millijoule, 2-line source with tunable frequency separation and pulse duration tailored for extending conversion efficiencies to those predicted by simulation. We perform difference frequency generation (DFG) experiments and demonstrate record conversion efficiencies near 1% at a frequency of 530 GHz, which is a factor of 3 beyond the current state of the art.

## 2. EXPERIMENTAL SETUP

Optical pulses were provided by a home-built laser consisting of a front-end [15], a commercial Yb:KYW regenerative amplifier and a Yb:YAG four-pass amplifier (4PA). The front-end was composed of two, commercial, single-frequency CW-lasers, one provided by Stable Laser Systems Inc., centered at 1.03 μm and frequency stabilized to 1 Hz, and the other by Toptica Photonics AG., tunable from 990 – 1080 nm, with specified linewidth of 1 MHz over 5 µs and temperature stability of 0.4 GHz/K, to control the wavelength separation for phase-matching in the PPLN crystal. The CW-lasers were combined in a polarization-maintaining fiber and then amplified and temporally chopped in successive steps using ytterbium-doped fiber amplifiers and acousto- and electro-optic modulators (EOM), respectively. The pulse duration and profile were thus determined by the bandwidth of the EOM, which produced flat-top pulses with rise times of ~ 60 ps. The output pulses had duration tunable in increments of 250 ps (i.e., 250, 500, 750 ps) and energy up to 20 mJ at 10 Hz, resulting in pulses matching the requirements for efficient THz generation described by Ravi *et al.* [11,16]. The optical pulses were then sent into a set of z-cut MgO:PPLN crystals of varying length, poling period and aperture (Table 1). LiNbO$_3$ was chosen as the nonlinear material due to its high effective nonlinear optical coefficient $d_{eff} = 168$ pm/V [2], and 5% MgO doping was used to lessen photorefractive effects [20]. The crystals were mounted in a cryostat and cooled to liquid nitrogen temperature (T ~ 80 K) to minimize THz absorption [17,18].

Table 1. Dimensions of MgO:PPLN crystals tested.

| Designation | $X_{212}^{4x4x40}$ | $X_{400}^{4x4x40}$ | $X_{212}^{4x4x20}$ | $X_{400}^{4x4x20}$ | $X_{212}^{3x3x20}$ |
|---|---|---|---|---|---|
| Aperture (mm) | 4 x 4 | 4 x 4 | 4 x 4 | 4 x 4 | 3 x 3 |
| Length (mm) | 40 | 40 | 20 | 20 | 20 |
| $\Lambda_{PPLN}$ (µm) | 212 | 400 | 212 | 400 | 400 |

The experimental setup, as well as the measured optical spectrum and temporal profile from the 4PA output are depicted in Fig. 1. The amplified beam had a Gaussian spatial profile with a $1/e^2$ radius of 0.8 mm and 1.1 mm in x- and y-directions, respectively, ensuring a collimated beam over a range longer than the MgO:PPLN crystals.

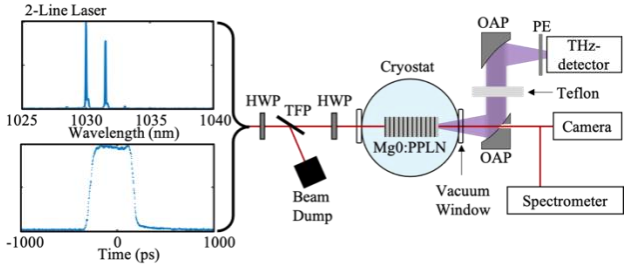

**Figure 1:** Experimental setup. HWP: half-wave plate, TFP: thin-film polarizer, OAP: off-axis parabolic mirror, PE: Polyethylene.

The pulse energy was tuned using a half-wave plate and thin-film polarizer, and the optical polarization was aligned with the crystal extraordinary c-axis to maximize the effective nonlinear coefficient. The generated THz radiation was collected and imaged onto a pyroelectric detector (Gentec-EO, SDX-1152) using a pair of 2" diameter, 4" focal length off-axis parabolic (OAP) mirrors. A 1.6 mm thick Teflon plate and a 2 mm polyethylene plate were placed in front of the detector to reduce noise in the THz energy measurement from the transmitted optical beam as well as parasitic second harmonic generation of the pump beam. The first OAP, which collects and collimates the THz energy, had a circular hole of 3 mm diameter in the center to separate the optical beam for monitoring the crystal output surface for damage as well as the spectral reshaping of the optical beam due to THz generation.

## 3. RESULTS AND DISCUSSION

The parameters varied in the experiments were the frequency separation of the two spectral lines, the optical pulse energy and the pulse duration. The frequency separation, $\Delta \nu$, was tuned first to optimize the phase matching within each crystal (Fig 2a, b).

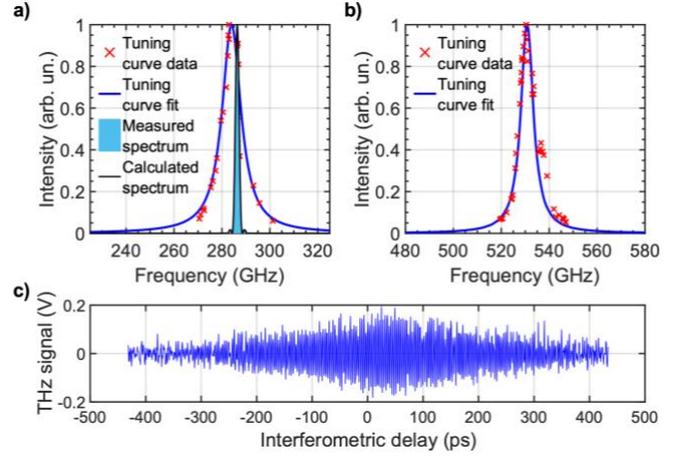

Figure 2: **a)** Tuning curve data and theoretical fit of crystal response function for $X_{400}^{4x4x40}$, together with the measured THz spectrum and calculation of expected THz spectrum. **b)** Tuning curve data and theoretical fit of crystal response function for $X_{212}^{4x4x20}$. **c)** THz time-domain interferogram for spectrum in a).

The measured THz yield vs $\Delta\nu$ effectively maps out the frequency response function of the crystal, which, following the approach in [13,19], is described in the undepleted-pump approximation by:

$$K(\Omega, L) \equiv \frac{\Omega^2 \chi^{(2)^2}_{eff}}{\frac{1}{4}\alpha^2(\Omega) + \Delta k^2(\Omega)} \left[ \left(1 - e^{-\frac{1}{2}\alpha(\Omega)L}\right)^2 + 4e^{-\frac{1}{2}\alpha(\Omega)L} \sin^2\left(\frac{\Delta k(\Omega) L}{2}\right) \right] \quad (1)$$

where $\Omega$ is the THz angular frequency, $\alpha(\Omega)$ is the THz absorption coefficient, $\chi^{(2)}_{eff}$ is the effective second-order nonlinear susceptibility, $L$ is the crystal length and $\Lambda_{PPLN}$ is the poling period. Furthermore, $\Delta k(\Omega) = \frac{2\pi}{\Lambda_{PPLN}} \frac{\Omega - \Omega_{PM}}{\Omega_{PM}}$ is the phase mismatch, while $\Omega_{PM} = \frac{2\pi c}{\Lambda_{PPLN}} \frac{1}{n_\phi(\Omega) - n_g(\omega)}$ is the phase-matched THz angular frequency. In the above, $n_\phi(\Omega)$ is THz index of refraction, $n_g(\omega) = c/v_g(\omega)$ is the optical group index, and $v_g(\omega)$ is the optical group velocity. In principle, comparison of $K(\Omega, L)$ with the data in Fig. 1 can be used for in-situ measurement of key material properties of LiNbO$_3$ in the THz domain which are still not universally agreed upon. Specifically, the THz index of refraction, $n_\phi(\Omega)$, determines the location of the tuning-curve peak, and $\alpha(\Omega)$ determines its width for crystals longer than the absorption length. Assuming $n_g(1.03 \ \mu m) = 2.2159$ [20], we find best agreement with the experimental results for $n_\phi(286 \text{ GHz}) = 4.885$, $\alpha(286 \text{ GHz}) = 4.5 \text{ cm}^{-1}$, $n_\phi(532 \text{ GHz}) = 4.880$ and $\alpha(532 \text{ GHz}) = 3.5 \text{ cm}^{-1}$. For these values of $\alpha$, the exponential and sinc-squared terms in Eq. 1 become negligible and the response function can be approximated by:

$$K(\Omega) \cong \frac{\Omega^2 \chi^{(2)^2}_{eff}}{\frac{1}{4}\alpha^2(\Omega) + \Delta k^2(\Omega)}. \quad (2)$$

The FWHM relative bandwidth of the tuning curve is thus approximated by $\Delta\Omega/\Omega_{PM} \approx \alpha(\Omega_{PM})\Lambda_{PPLN}/2\pi$, or 1.2% and 2.9% for the 212 µm and 400 µm period crystals respectively. These best-fit absorption factors are significantly higher than reported in the literature for cryogenic temperatures and do not have the expected dependence on frequency [21]. As the Toptica laser adjustment was manual and affected by mode hopping, we conclude the tuning curve measurement was not sufficiently accurate for an absorption-coefficient determination. Nevertheless, Fig. 2 shows that the optimum wavelength separations are as expected and that the two-line approach successfully generates tunable monochromatic THz.

The spectrum of the THz emitted from the 400 µm period crystal (Fig. 2a), which was measured using time-domain interferometry (Fig. 2c), also matched well with expectations. The relative spectral width of 0.5% is significantly narrower than the crystal response function, indicating that the terahertz spectrum was defined by the properties of the optical pulses which were transform-limited with linewidths of 2.5 – 7.5 pm. The THz power spectrum can be calculated using: $I(\Omega) \propto |R(\Omega)|^2$, where $R(\Omega) \equiv \int_{-\infty}^{\infty} A_{op}(\omega + \Omega)A_{op}^*(\omega)d\omega$ and $A_{op}(\omega)$ is the spectral field of the driving optical pulse. To estimate $A_{op}(\omega)$, the temporal profile of the optical pulse (Fig. 1) was measured with a sampling oscilloscope, and fit to a supergaussian function of width 500 ps and order 6. As the spectral lines were too narrow to be accurately measured with our spectrometer, two transform-limited spectral lines centered at the peaks of the measured optical spectra were assumed. The calculated and measured THz spectra match nearly perfectly, including the side bands due to the flat-top temporal profile, see Fig. 2a.

Once the frequency separation was optimized, the THz yield was characterized as a function of optical pulse energy. The metric used for the yield is the "internal" conversion efficiency (CE), i.e., the ratio of THz to optical pulse energy within the crystal. The *internal* CE quantifies the effectiveness of the intrinsic THz generation process decoupled as much as possible from the practical issues of beam input and output coupling and THz transport. The internal THz pulse energy is inferred from the measured energy by correcting for losses from Fresnel reflections at the uncoated crystal and cryostat interfaces as well as from absorption in the optical attenuators. These losses were determined using calculation and characterization with a THz time-domain spectrometer (Table 2). The optical pulse internal energy was inferred similarly.

Fig. 3 shows measurements of the CE as a function of optical peak fluence and intensity for pulse durations of 250 ps, 500 ps and 750 ps. As expected, for a given fluence, the shorter pulses provide a higher THz yield due to the higher peak intensities. However, plotting against intensity shows equivalent performance independent of pulse length. The graphs show that despite the onset of saturation of the conversion process, it may be possible to increase yield by going to higher fluences and intensities. However, we limited our scans to a peak fluence of ~500 mJ/cm², (pulse energy of 7.3 mJ), to minimize the probability of photorefractive-induced damage.

For comparison, we ran 1D numerical simulations which include the effects of pump depletion as well as self-phase modulation. As the tuning curve analysis above was inconclusive, values of the absorption coefficient for the simulations of $\alpha(532\text{ GHz}) = 1.1\text{ cm}^{-1}$ and $\alpha(286\text{ GHz}) = 0.73\text{ cm}^{-1}$ were taken from the literature [21]. The resulting simulations agree quantitatively with the measured CEs within about 20% which is within a range attributable to tuning variations of the highly-sensitive spatial and spectral alignments as well as in the temperature which affects the THz absorption. The behavior with fluence and pulse duration are also well captured. In Fig. 3b, the simulations predict a slight advantage for the longer pulses, which is expected due to better matching between the lengths of the optical pulse and the crystal [11]. In Fig. 3a, analytic calculations using the formalism in [11] were also done to cross-check the validity of the simulations, and show perfect agreement at low fluence where the undepleted approximation is valid.

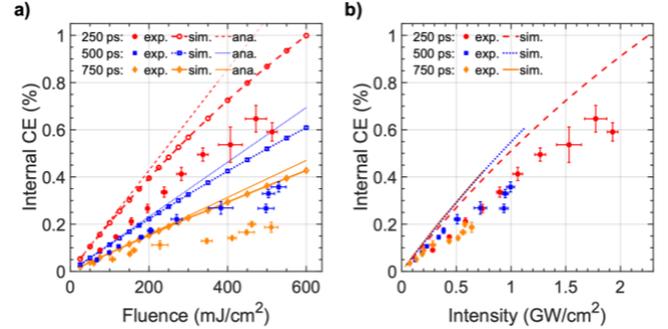

Fig 3. a) Experimental data, simulation and analytic calculation of CE vs peak fluence for crystal $X_{212}^{3x3x20}$ for pulse durations of 250 ps (blue), 500 ps (red) and 750 ps (yellow) at cryogenic temperature. b) Data and simulation from a) plotted vs intensity.

Table 2. THz losses in various optical elements

| Frequency (GHz) | Teflon plate (%) | Cryostat window (%) | Polyethylene plate (%) |
|---|---|---|---|
| 286 | 6.6 | 34.4 | 31.7 |
| 530 | 6.7 | 47.8 | 47.1 |

To evaluate the effects of crystal length and poling period, measurements of THz yield vs pulse energy and corresponding simulations were performed for the five crystals in Table 1 (Fig. 4). These data were taken for the shortest pulse duration of 250 ps in order to maximize the yield. A record conversion efficiency of 0.9% was achieved for crystal $X_{212}^{4x4x40}$. For both 4 cm crystals, the agreement between data and simulation is excellent, including the onset of saturation at higher fluence.

According to both the simulations and the data, the 212 µm poled crystals provide a higher CE than the 400 µm poled crystals. This effect can be understood analytically. From Eq. 2, for optimized phase matching (i.e., $\Delta k = 0$), the CE should scale roughly as: $CE(\Omega) \propto \Omega^2/\alpha^2(\Omega)$. Using our values of frequency and absorption factor, we should expect $CE(532\text{ GHz}) \approx 1.5 \times CE(286\text{ GHz})$, which approximately matches the data in the lower fluence range. Strictly speaking, Eqs. 1 and 2 only apply when the pump laser is completely unaffected by the conversion process, which is best satisfied at low fluences. At higher fluences, the simulations and data both clearly show that saturation, which is related to the pump evolution, occurs earlier for the 212 µm poling case. Similar efficiencies may therefore be reachable with 400 µm poling at higher fluence, provided the crystals do not damage.

The simulation and data also agree that the 2 cm long crystals provide a lower CEs than the 4 cm counter parts. We note that crystals $X_{212}^{4x4x20}$ and $X_{400}^{4x4x20}$, which were used extensively in the testing phase of the experiments, showed significant signs of optical damage. Crystal $X_{212}^{3x3x20}$, which was pristine for these measurements, yielded a CE of approximately 1.6x greater than crystal $X_{212}^{4x4x20}$, thus providing an estimate of the damage level. For this reason, crystal $X_{212}^{3x3x20}$ is the most comparable to the 4 cm crystals, which were also lightly used, despite the difference in aperture size. Unfortunately, an un-used 2 cm, 400 μm poled crystal was not available for direct comparison.

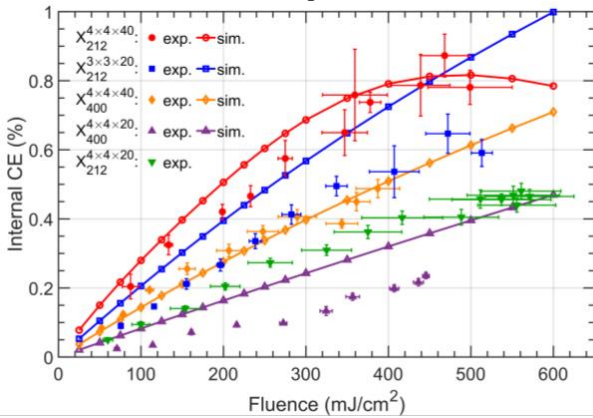

**Figure 4.** Internal CE vs laser fluence for the fixed pulse duration of 250 ps for crystals of varying poling period and length (dots: experiment, circles: simulation).

The role of cascading in achieving high efficiencies implies a strong impact of the interaction on the output optical spectrum. Indeed, the measured spectrum shows the appearance of additional distinct lines (primarily on the long-wavelength side) increasing in number with fluence to over 13 at the maximum (Fig. 5). The shift of the spectral center of mass by over 10 nm to long wavelengths thus confirms percent-level optical conversion to THz [13].

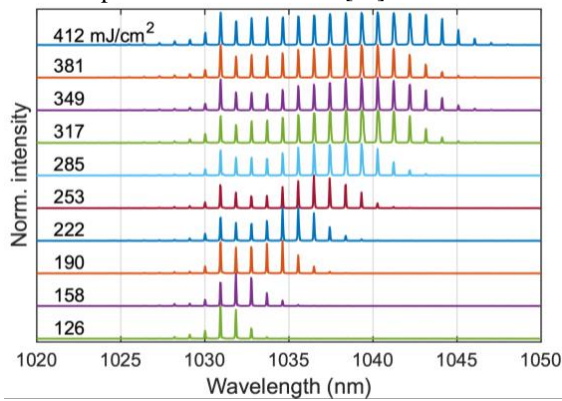

Fig. 5. Measured, individually normalized optical spectra after THz generation for varying optical input energies in crystal $X_{400}^{4x4x40}$.

## 4. CONCLUSION

We investigated narrow-band THz generation using a customized two-spectral-line laser source in cryogenically-cooled MgO:PPLN crystals, and achieved conversion efficiencies of 0.9% for 0.53 THz and 0.5% for 0.29 THz which are a factor of 3 above previous reports [9]. These results confirm the benefits of tailoring the laser source [11] for nonlinear optical conversion to narrowband THz and point to even higher efficiencies [19] enabling mJ-scale THz applications.

**Funding.** This work was supported by project KA908-12/1 of the Deutsche Forschungsgemeinschaft and by the European Research Council under the European Union's Seventh Framework Program (FP7/2007-2013) through Synergy Grant AXSIS (609920). Dr. Tian acknowledges support from the Helmholtz – OCPC Postdoc Program. **Disclosures**. The authors declare no conflicts of interest.

## References

1. J. P. Guillet, B. Recur, L. Frederique, B. Bousquet, L. Canioni, I Manek-Hönninger, P. Desbarats and P. Mounaix, "Review of terahertz tomography techniques. Journal of Infrared, Millimeter, and Terahertz Waves," J Infrared Milli Terahz Waves 35, 382–411 (2014).
2. J. Hebling, K.-L. Yeh, M. C. Hoffmann, and K. A. Nelson, "High-power THz generation, THz nonlinear optics and THz nonlinear spectroscopy," IEEE J. Sel. Top. Quantum Electron. 14, 345–353 (2008).
3. M. Beck, I. Rousseau, M. Klammer, P. Leiderer, M. Mittendorff, S. Winnerl, M. Helm, G. N. Gol'tsman, and J. Demsar, Phys. Rev. Lett. 110, 267003 (2013).
4. E. A. Nanni, W. R. Huang, K.-H. Hong, K. Ravi, A. Fallahi, G. Moriena, R. J. D. Miller, and F. X. Kärtner, "Terahertz-driven linear electron acceleration," 6, 8486 (2015).
5. D. Zhang, A. Fallahi, M. Hemmer, X. Wu, M. Fakhari, Y. Hua, H. Cankaya, A.-L. Calendron, L. E. Zapata, N. H. Matlis, and F. X. Kärtner, "Segmented terahertz electron accelerator and manipulator (STEAM)," 12, 336–342 (2018).
6. C. Vicario, B. Monoszlai, and C. P. Hauri, "GV/m single-cycle terahertz fields from a laser-driven large-size partitioned organic crystal," Phys. Rev. Lett. **112**, 213901 (2014).
7. S. Carbajo, J. Schulte, X. Wu, K. Ravi, D. N. Schimpf, and F. X. Kärtner, "Efficient narrowband terahertz generation in cryogenically cooled periodically poled lithium niobate," Opt. Lett. 40, 5762–5765 (2015).
8. F. Ahr, S. W. Jolly, N. H. Matlis, S. Carbajo, T. Kroh, K. Ravi, D. N. Schimpf, J. Schulte, H. Ishizuki, T. Taira, A. R. Maier, and F. X. Kärtner, "Narrowband terahertz generation with chirped-and-delayed laser pulses in periodically poled lithium niobate," Opt. Lett. 42, 2118–2121 (2017).
9. S. W. Jolly, N. H. Matlis, F. Ahr, V. Leroux, T. Eichner, A.-L. Calendron, H. Ishizuki, T. Taira, F. X. Kärtner, and A. R. Maier, "Spectral phase control of interfering chirped pulses for high-energy narrowband terahertz generation," Nature Comm. 10, 2591 (2019).
10. F. Lemery, T. Vinatier, F. Mayet, R. Aßmann, E. Baynard, J. Demailly, U. Dorda, B. Lucas, A.-K. Pandey and M. Pittman, "Highly scalable multicycle THz production with a homemade periodically poled macrocrystal." Commun Phys 3, 150 (2020).
11. K. Ravi, D.N Schimpf and F. X Kärtner, "Pulse sequences for efficient multi-cycle terahertz generation in periodically poled lithium niobate", Opt. Express, **24**, 25582-25607 (2016).
12. M. Cronin-Golomb, "Cascaded nonlinear difference-frequency generation of enhanced terahertz wave production," Opt. Lett. **29**, 2046–2048 (2004).
13. K. L. Vodopyanov, "Optical generation of narrow-band terahertz packets in periodically-inverted electro-optic crystals: conversion efficiency and optimal laser pulse format," Opt. Express **14**, 2263–2276 (2006).
14. M. Lang and A. Deninger, "Laser-based terahertz generation & applications," Photonik International (2012).
15. D. N. Schimpf, H. T. Olgun, A. Kalaydzhyan, Y. Hua, N. H. Matlis, and F. X. Kärtner, "Frequency-comb-based laser system producing stable optical beat pulses with picosecond durations suitable for high-precision multi-cycle terahertz-wave generation and rapid detection," Opt. Express **27**, 11037–11056 (2019).
16. L. Wang, A. Fallahi, K. Ravi, and F. Kärtner, "High efficiency terahertz generation in a multi-stage system," Opt. Express 26, 29744–29768 (2018).
17. J.-I. Shikata, M. Sato, T. Taniuchi, H. Ito, and K. Kawase, "Enhancement of terahertz-wave output from LiNbO3 optical parametric oscillators by cryogenic cooling," Opt. Lett. **24**, 202–204 (1999).
18. S.-W. Huang, E. Granados, W. R. Huang, K.-H. Hong, L. E. Zapata, and F. X. Kärtner, "High conversion efficiency, high energy terahertz pulses by optical rectification in cryogenically cooled lithium niobate," Opt. Lett. **38**, 796–798 (2013).
19. K. Ravi and F. X. Kärtner, "Raman Shifting Induced by Cascaded Quadratic Nonlinearities for Terahertz Generation," Laser Photon. Rev. 14, 2000109 (2020).
20. D. E. Zelmon, D. L. Small, and D. Jundt. "Infrared corrected Sellmeier coefficients for congruently grown lithium niobate and 5 mol.% magnesium oxide-doped lithium niobate", J. Opt. Soc. Am. B **14**, 3319-3322 (1997).
21. J. A. Fülöp, L. Pálfalvi, M. C. Hoffmann and J. Hebling, "Towards generation of mJ-level ultrashort THz pulses by optical rectification," Opt. Express **19**, 15090 – 15097 (2011).